\documentclass[preprint,showpacs,preprintnumbers,nofootinbib,amsmath,amssymb,aps,prc,longbibliography]{revtex4-1}
\usepackage{graphicx}
\usepackage{dcolumn}
\usepackage{bm}
\begin{document}
\title{Giant dipole resonance in $^{88}$Mo from
phonon damping model's strength functions 
averaged over temperature and angular momentum distributions}
\author{N. Dinh Dang$^{1}$}
  \email{dang@riken.jp}
  \author{M. Ciemala$^{2}$}
 \email{michal.ciemala@ifj.edu.pl}
  \author{M. Kmiecik$^{2}$}
 \email{Maria.Kmiecik@ifj.edu.pl}
\author{A. Maj$^{2}$}
 \email{Adam.Maj@ifj.edu.pl}
 \affiliation{1) Theoretical Nuclear Physics Laboratory, RIKEN Nishina Center
for Accelerator-Based Science,
2-1 Hirosawa, Wako City, 351-0198 Saitama, Japan\\
and Institute for Nuclear Science and Technique, Hanoi, Vietnam\\
2) Niewodniczanski Institute of Nuclear Physics PAN, 31-342 Krak\'{o}w, Poland}

\date{\today}
\begin{abstract}
The line shapes of giant dipole resonance (GDR) in the decay of the compound nucleus $^{88}$Mo, which is formed after the fusion-evaporation reaction $^{48}$Ti $+$ $^{40}$Ca at various excitation energies $E^{*}$ from 58 to 308 MeV, are generated by averaging the GDR strength functions predicted within the phonon damping model (PDM) using the empirical probabilities for temperature  and angular momentum. The average strength functions are compared with the PDM strength functions calculated at the mean temperature  and mean angular momentum, which are obtained by averaging the values of temperature and angular momentum using the same temperature and angular-momentum probability distributions, respectively. It is seen that these two ways of generating the GDR linear line shape yield very similar results. It is also shown that the GDR width approaches a saturation at angular momentum $J\geq$ 50$\hbar$ at $T=$ 4 MeV and at $J\geq$ 70$\hbar$ at any $T$.
\end{abstract}

                              
\maketitle
\section{Introduction}
\label{Intro}
Many theoretical and experimental studies in nuclear structure during the last three decades were devoted to the giant dipole resonance (GDR) in highly excited nuclei (see Ref.\cite{Gaa,Bortignon,Woude} for reviews of the subject). A recent compilation of the experimental systematics of the GDR built on excited states is given in Ref. \cite{Schiller}. The most recent experimental measurements are reported in Ref. \cite{Mo88,Mo88_2} for GDR in the fusion-evaporation reaction forming the compound nucleus $^{88}$Mo at high temperature and angular momentum, and in Ref. \cite{alpha}, where the GDR width in $^{201}$Tl in the $\alpha$ induced fusion reaction was extracted at low temperature. The center of attention has been the evolution of the GDR width as functions of temperature $T$ and angular momentum $J$. The GDR line shape and its full-width at half maximum (FWHM) $\Gamma_{GDR}$  are experimentally extracted  from the  statistical calculations, which use the Lorentzian strength function to reproduce the $\gamma$-ray spectra detected from the decay of the highly-excited compound nucleus at the excitation energy $E^*$. They are often compared with the theoretical predictions, which are obtained at a given values of $T$ and/or $J$. 

The extraction of nuclear temperature $T$ and angular momentum $J$ is crucial for a meaningful comparison between experiment and theory because the initial temperature $T_{max}$ and/or angular momentum $J_{max}$ at the first step in the decay of the compound nucleus are significantly higher than the mean values $\overline{T}$ and  $\overline{J}$, obtained by averaging over all daughter nuclei in the decay process. Moreover, while the theoretical GDR strength function is calculated at a fixed value of $T$ and/or $J$, its experimental counterpart is extracted by fitting the spectrum, which is generated by a multistep cascade decay, where the nucleus undergoes a cooling down from the initial maximal value of $T_{max}$ (and/or $J_{max}$). Because of this mechanism, the authors of Ref. \cite{Gervais} have proposed to incorporate the theoretical strength functions into the full statistical decay calculations and compare the results obtained with the experimental data. This method was applied to test the validity of several theoretical models in Refs. \cite{Gervais,Eisenman}, namely the collisional damping model (CDM)~\cite{CDM}, the thermal shape fluctuation model (TSFM)~\cite{TSFM}, and the phonon damping model (PDM)~\cite{PDM1,PDM2,PDM3}. The CDM studies the GDR evolution within the framework of the macroscopic Landau-Vlasov theory that includes the collision term in the Landau integral. The TSFM describes the GDR line shape by calculating the GDR cross section as a thermal average over all shape-dependent cross sections under quadrupole deformations.  The PDM describes the broadening of the GDR width at low and medium $T$ and $J$ as well as its saturation at high $T$ and $J$ via coupling of the GDR to noncollective particle-hole (ph), particle-particle (pp) and hole-hole (hh) configurations. The detailed analysis in Ref. \cite{Gervais} shows that neither the TSFM nor CDM could reproduce the GDR data for $^{120}$Sn, whereas Ref. \cite{Eisenman} demonstrates that the PDM describes reasonably well the GDR line shape at $\overline{T}\geq$ 2 MeV.  By including the nonvanishing thermal pairing gap, the PDM is also capable of correctly describing the temperature dependence of the GDR width at low temperature ($T<$ 2 MeV)~\cite{PDM2,PDM4}.  However, a question still remains open, namely it is not clear if the GDR line shape obtained by averaging the GDR strength functions in the whole interval of $T$ and/or $J$, within which the daughter nuclei are populated, is equivalent to the GDR strength function obtained at the mean values $\overline{T}$ of temperature and $\overline{J}$ of angular momentum in these intervals. Resolving this issue has a practical importance since if the answer is positive, one can avoid the calculations of many strength functions as the temperature and/or angular momentum decreases starting from $T_{max}$ and/or  $J_{max}$ to obtain the average line shape, and use the strength function obtained at one given pair of values $\overline{T}$ and $\overline{J}$ instead. The aim of the present paper is to answer this question.

The paper is organized as follows. The formalism is presented in Sec. \ref{formalism}. The results of numerical calculations for GDR strength functions within the PDM, which are averaged by using the empirical probabilities distributions for temperature and angular momentum in $^{88}$Mo at various excitation energies $E^{*}$, are discussed in Sec.  \ref{results}. The paper is summarized in the last section, where conclusions are drawn.
\section{Formalism}
\label{formalism}
\label{PDM}
\subsection{Model Hamiltonian and GDR strength function at finite temperature and angular momentum}
The formalism of the PDM, whose Hamiltonian describes a hot spherical system noncollectively rotating
about the symmetry $z$-axis, has been presented and discussed thoroughly in Ref. \cite{PDM3}.
Therefore we summarize here only the final results, which are necessary for the analysis in the present paper. 

The model Hamiltonian is given as 
\begin{equation}
H=H_{0} - \gamma\hat{M}~,
\label{H0}
\end{equation}
where $H_0$ is the PDM Hamiltonian of the non-rotating system, described in Ref. \cite{PDM1,PDM2}, and $\hat{M}$ represents the total angular momentum $\hat{J}$, which, in the present case, coincides with its $z$-projection $\hat{M}$, that is  
\begin{equation}
\hat{J}\equiv\hat{M} =\sum_{k>0}m_{k}(N_{k}-N_{-k})~.
\label{M}
\end{equation}
Here the subscripts $k$ denote the single-particle states 
$|k,m_{k}\rangle$ in the deformed basis with the angular momentum $k$ and the positive single-particle spin
projection $m_{k}$, whereas the subscripts $-k$  denote the
time-reversal states $|k,-m_{k}\rangle$ ($m_{k}>$ 0). The particle number operator $\hat{N}$ is written as
\begin{equation}
\hat{N} = \sum_{k>0}(N_{k}+N_{-k})~,\hspace{5mm} N_{\pm k}=a_{\pm k}^{\dagger}a_{\pm k}~,
\label{N}
\end{equation}
where $a_{\pm k}^{\dagger}$
($a_{\pm k}$) is the creation (annihilation) operator of a particle with spin $k$, spin projection $\pm m_k$, 
and energy $\epsilon_{k}$. 
By using Eqs. (\ref{M}) and (\ref{N}), the Hamiltonian (\ref{H0}) transforms into
\[
H = \sum_{k>0}(\epsilon_k-\lambda
-\gamma m_k)N_k+\sum_{k>0}(\epsilon_k-\lambda+\gamma m_k)N_{-k}+
\sum_{q}\omega_{q}Q_{q}^{\dagger}Q_{q}
\]
\begin{equation}
+\sum_{k,k'>0}\sum_q{\cal F}_{kk'}^{(q)}(a_{k}^{\dagger}a_{k'}+a_{-k}^{\dagger}a_{-k'})(Q_{q}^{\dagger}+Q_{q})~,
\label{H}
\end{equation}
where $\lambda$ denotes the chemical potential.
The particle (p) states correspond to those with $\epsilon_k>\lambda$, whereas the hole (h) states are those with $\epsilon_k<\lambda$. The operator $Q_{q}^{\dagger}$ ($Q_{q}$) is the phonon creation (annihilation) operator for a collective vibration with energy $\omega_{q}$. 
In this way, Hamiltonian (\ref{H}) describes two mean fields, the single-particle mean field (the first two terms on the right-hand side of Eq. (\ref{H})),  and the phonon one associated with the GDR (the third term), as well as the coupling between them (the last term) with matrix elements ${\cal F}_{kk'}^{(q)}$. The GDR acquires a width and the phonon energy $\omega_q$ undergoes a shift because of this coupling. 
By including the angular momentum in the first two terms, each of spherical orbital $j$ with energy $\epsilon_j$ splits into $2\Omega_j=2j + 1$ distinctive levels, half of which consists of levels with energies $\epsilon_k+\gamma m_k$ , whereas the other half consists of levels with energies $\epsilon_k-\gamma m_k$, with $k = 1, ..., \Omega/2$, where $\Omega=2\sum_j\Omega_j$ is the total number of levels. Because the effect of thermal pairing on the GDR width is negligible in the region of moderate (high) $T$ and $J$ ($E{^*}\geq$ 58 MeV for $^{88}$Mo~\cite{PDM2,PDM4}, we neglect it it in the calculation of the GDR strength functions for simplicity. 

The chemical potential $\lambda$ and the rotation frequency $\gamma$ are defined from the equations for conservation of the angular momentum $J$ and particle number $N$ as
\begin{equation}
J = \sum_k m_k(f_{k}^{+}-f_{k}^{-})~,\hspace{5mm} N =\sum_k(f_{k}^{+}+f_{k}^{-})~,
\label{M&N}
\end{equation}
where $J=\langle\hat{J}\rangle=M=\langle\hat{M}\rangle$, $N=\langle\hat{N}\rangle$, $f_{k}^{\pm} =
\langle N_{\pm k}\rangle$ with the grand canonical ensemble average $\langle...\rangle\equiv{\rm Tr}[...{\exp}(-\beta H)]/{\rm Tr}[{\exp}(-\beta H)]$  ($\beta = T^{-1}$). The single-particle occupation numbers $f_k^{\pm}$ are approximated with the Fermi-Dirac distribution: 
\begin{equation}
f_k^{\pm} =\frac{1}{\exp(\beta E_k^{\mp})+1}~,\hspace{5mm} E_k^{\mp} = \epsilon_k-\lambda\mp\gamma m_k~.
\label{nk}
\end{equation}

By using the Hamiltonian (\ref{H}) and the method of double-time Green's functions, the final equation for the Green's function, which describes the phonon propagation, was derived in Ref. \cite{PDM3} as
\begin{equation}
G_q(E) = \frac{1}{2\pi}\frac{1}{E-\tilde\omega_q}~,\hspace{2mm} \tilde\omega=\omega_q+P_q(E)~,
\hspace{2mm} P_q(E) = \sum_{kk'}[{\cal F}_{kk'}^{(q)}]^{2}\bigg[\frac{f^{+}_{k'}-f^{+}_{k}}
{E-E_k^{-}+E_{k'}^{-}}+\frac{f^{-}_{k'}-f^{-}_{k}}
{E-E_k^{+}+E_{k'}^{+}} \bigg]~.
\label{Gfinal}
\end{equation}
The principal value of the polarization operator $P_q(\omega)$ at a real $\omega$ defines the energy shift from the unperturbed phonon energy $\omega_q$ to $\tilde\omega_q$ under the effect of particle-phonon coupling, whereas the 
imaginary part $\gamma_q(\omega) = \Im mP_q(\omega\pm i\varepsilon)$ of the analytic continuation of $P_q(E)$ into the complex energy plan $E=\omega\pm  i\varepsilon$ defines the phonon damping 
$\gamma_q(\omega)$, whose final explicit expression reads
\begin{equation}
\gamma_q(\omega) = \varepsilon\sum_{kk'}[{\cal F}_{kk'}^{(q)}]^{2}
\bigg[\frac{f_{k'}^{+}-f_{k}^{+}}
{(\omega-E_k^{-} + E_{k'}^{-})^2+\varepsilon^2}+\frac{f_{k'}^{-}-f_{k}^{-}}{(\omega-E_k^{+} + E_{k'}^{+})^{2}+\varepsilon^{2}}\bigg]~,
\label{gamma1}
\end{equation}
where the representation $\delta(x) =\lim_{\varepsilon\rightarrow 0}\varepsilon/[\pi(x^{2}+\varepsilon^2)]$ is used to smooth the $\delta$-functions and to effectively take into account the contribution of the escape width owing to the coupling to continuum. 

The spectral intensity is found from the analytic properties of Green's function (\ref{Gfinal}) as ${\cal J}_q(\omega) = i[G_q(\omega+i\varepsilon) - G_q(\omega-i\varepsilon)]/[{\rm e}^{\beta\omega}-1]$, from which one obtains the GDR strength function $S(\omega)$ 
as $S(\omega) = \widetilde{\cal J}_q(\omega)[\exp(\beta\omega)-1]$, where $\widetilde{\cal J}_q(\omega)$ 
denotes ${\cal J}_q(\omega)$ calculated at  the GDR energy $\tilde\omega_q=E_{GDR}$~\cite{PDM1,PDM2,PDM3}.
The final result is a Breit-Wigner-like distribution with the energy-dependent half width $\gamma_q(\omega)$:
\begin{equation}
S_{BW}(\omega) =
\frac{1}{\pi}\frac{\gamma_q(\omega)}{[\omega-E_{GDR}]^2+\gamma_q^2(\omega)}~.
\label{SBW}
\end{equation}
The FWHM $\Gamma(T)$ of the GDR 
is defined as a function of $T$ and $J$ as~\cite{PDM1,PDM2}
\begin{equation}
\Gamma(T,J) = 2\gamma_q[\omega=E_{GDR}]~.
\label{FWHM}
\end{equation}
The evaporation width of the compound nuclear states, which comes from the quantum mechanical uncertainty principle~\cite{evaporation} is not included because its effect on the GDR width is expected to be significant only at large values of the average temperature ($\gg$ 3.3 MeV) and average angular momentum ($\gg$ 30$\hbar$)~\cite{Gervais}.
For the comparison with the experimental line shape, which is fitted by using the Lorentzian distribution, it is convenient to use 
the following Lorentzian-like strength function~\cite{Lor},  which is composed of two Breit-Wigner-like distributions (\ref{SBW}) 
(see, e.g., Eq. (16) of Ref. \cite{viscosity}):
\begin{equation}
S_L(\omega) =
\frac{\omega}{E_{GDR}}[S_{BW}(\omega,E_{GDR})-S_{BW}(\omega,-E_{GDR})]~.
\label{SL}
\end{equation}
\subsection{Averaging over the probability distributions of temperature and angular momentum}
In the fusion-evaporation reactions, where two heavy nuclei coalesce at high energy $E^{*}$ far above the Coulomb barrier, the resulting compound system at high angular momentum decays by evaporating particles in competition with high-energy $\gamma$ rays. The GDR is extracted from the high-energy $\gamma$ ray spectrum as a Lorentzian located at energy of around  
$E_{GDR} = 17 A^{-1/3} + 25 A^{-1/6}$\cite{Gaa}. During this $\gamma$ ray emission the nuclear temperature decreases from its initial value $T_m$, resulting in a probability distribution $p_T(T_i)$ of temperature, where $T_i$ is the temperature of the $i$-th step in the statistical decay. The same takes place with the angular momentum, which decreases from its initial value $J_n$, resulting in the  probability distribution $p_J(J_j)$.

Given the temperature and angular momentum probabilities distributions $p_T(T_i)$ and $p_J(J_i)$, the average strength function at the excitation energy $E^{*}$ is calculated as
\begin{equation}
S_{k}(\omega,E^{*}) =\frac{\sum_ip_J(J_i)\overline{S}_k(\omega,J_i)}{\sum_ip_J(J_i)}~,\hspace{5mm} \overline{S}_k(\omega,J) = \frac{\sum_jp_T(T_j)S_k(\omega,T_i,J)}{\sum_jp_T(T_j)}~,
\label{averS}
\end{equation}
where the strength function $S_k(\omega,T, J)$ can be either $S_{BW}(\omega)$ (\ref{SBW}) ($k=BW$) or $S_L(\omega)$ (\ref{SL}) ($k=L$) obtained at a given pair of values $(T,J) = (T_i, J_i)$, whereas $\overline{S}_k(\omega,J)$ is the strength function obtained by averaging   $S_k(\omega,T, J)$ over the probability distribution of temperature at each value  $J_i$ of the angular momentum.
The average temperature $\overline{T}$ within the interval $T_1\leq T_i \leq T_m$, where the probability distribution $p_T(T_i)$ is determined 
($i = 1, 2, ...,m$), and the average angular momentum $\overline{J}$ within the interval $J_1\leq J_i \leq J_n$, where the probability distribution $p_M(M_j)$ is determined ($j = 1, 2, ...,n$), are calculated as 
\begin{equation}
\overline{T} = \frac{\sum_{i=1}^{m}p_T(T_i)T_i}{\sum_{i=1}^{m}p_T(T_i)}~,\hspace{5mm}  \overline{J} = \frac{\sum_{j=1}^{n}p_J(J_i)J_j}{\sum_{j=1}^{n}p_J(J_j)}~.
\label{averTM}
\end{equation}
In the present paper the average strength function $S_L(\omega, E^*)$ in Eq. (\ref{averS}) will be compared with the strength function $S_L(\omega,\overline{T},\overline{J})$
in Eq. (\ref{SL}), which is calculated at the average temperature $\overline{T}$ and average angular momentum $\overline{J}$ in Eq. (\ref{averTM}).
\section{Analysis of numerical results}
\label{results}
    \begin{figure}
       \includegraphics[width=9cm]{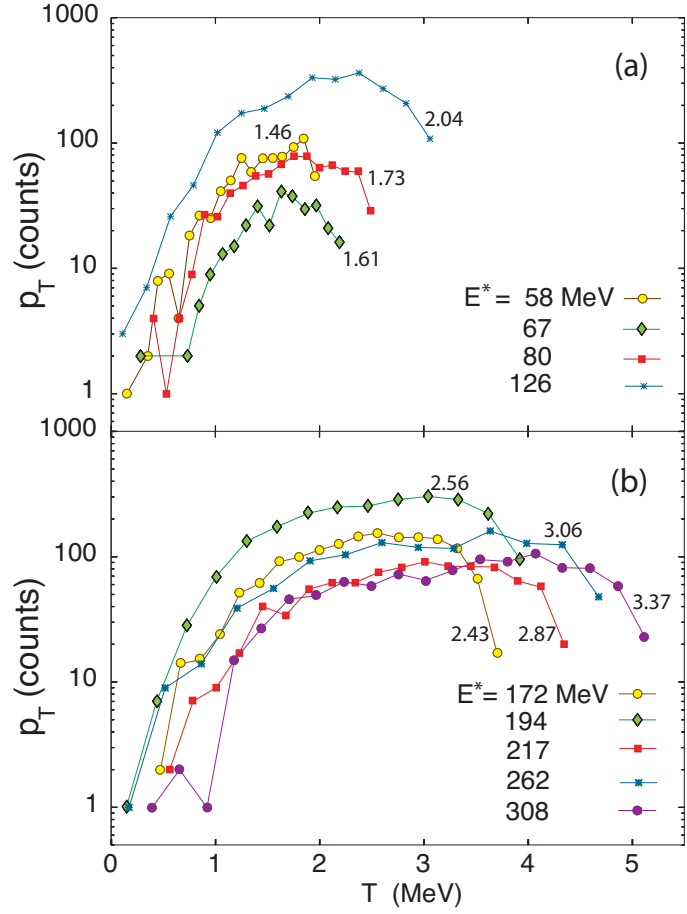}
       \caption{(Color online) Temperature distributions for $^{88}$Mo at excitation energies $E^{*}=$ 58, 67, 80, and 126 MeV (a) and 172, 194, 217, 262, and 308 MeV (b), which are calculated by the {\scriptsize GEMINI++} code (see text for details). The lines are drawn to guide the eye. The value of average temperature $\overline{T}$ (in MeV) at each energy is shown as a number at the corresponding line.
        \label{Tdis}}
    \end{figure}
    \begin{figure}
       \includegraphics[width=12cm]{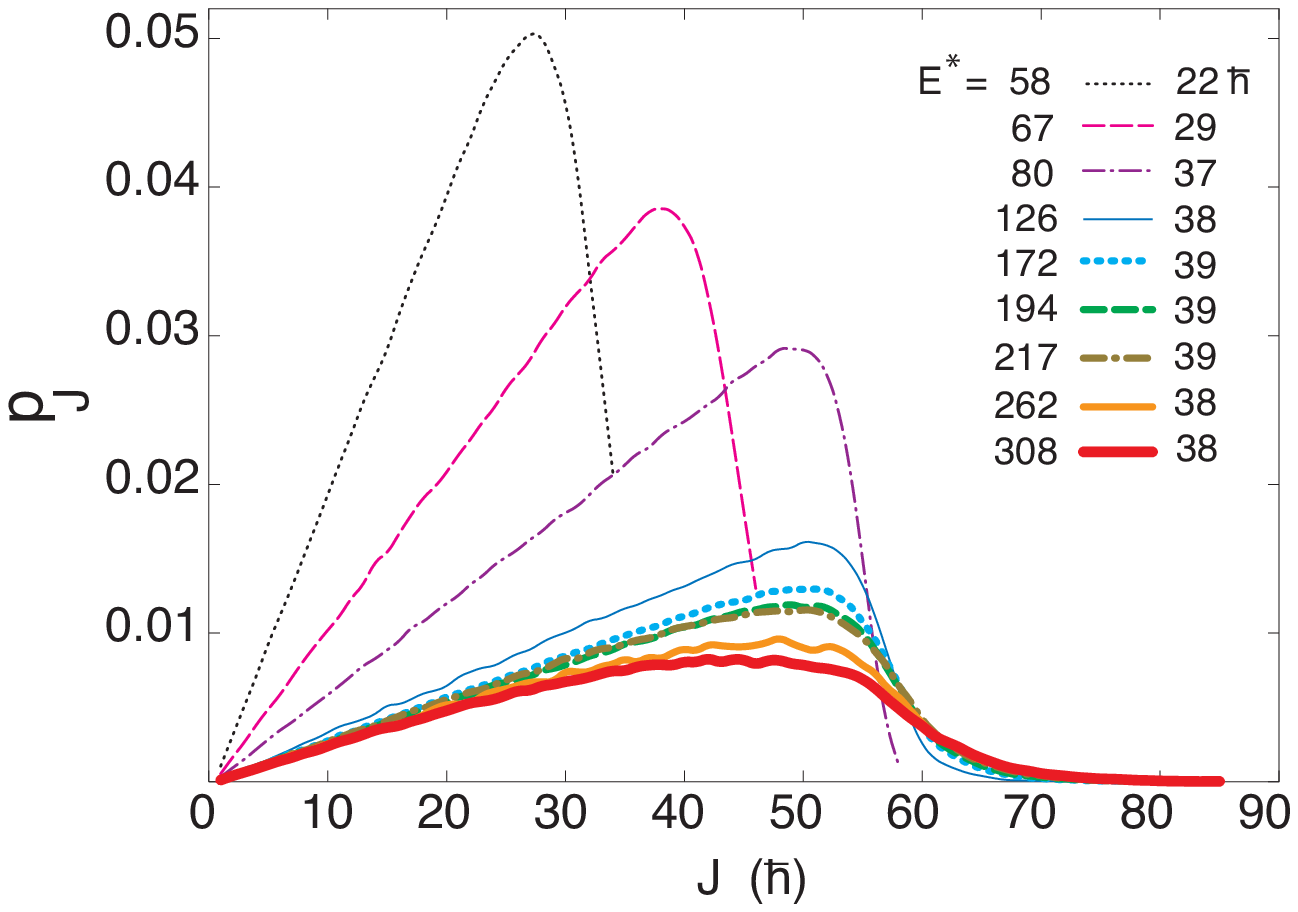}
       \caption{(Color online) Angular-momentum distributions for $^{88}$Mo at different excitation energies $E^{*}$, which are calculated by the {\scriptsize GEMINI++} code (see text for details), as shown in the panel (in MeV) with the corresponding average angular momenta $\overline{J}$ (in $\hbar$). 
               \label{Jdis}}
    \end{figure}
\subsection{Ingredients of the numerical calculations}
We employ the single-particle energies $\epsilon_{k}$, which are obtained from the Woods-Saxon potentials for neutrons and protons in $^{88}$Mo.  They span a large space at $J=$ 0 starting from the bottom $1s_{1/2}$ level located at around $-40$ MeV for neutrons and $-30$ MeV for protons up to around 22 MeV, where the part of the spectrum with positive values $\epsilon_k$ simulates an effective discretized continuum. These single particle energies are assumed to be temperature-independent based on the estimation within the temperature-dependent self-consistent Hartree-Fock calculations~\cite{Quentin}, which have demonstrated that the single-particle energies in heavy nuclei weakly change with $T$ up to $T\sim$ 5 MeV.

The matrix elements ${\cal F}_{ph}^{(q)}$ for the coupling of the GDR to non-collective $ph$ configurations, causing the quantal width already at $T=$ 0, are assumed to be the same and equal to ${\cal F}_1$,  whereas those for the coupling of the GDR to $pp$ ($hh$) configurations, ${\cal F}_{pp}^{(q)}$ and ${\cal F}_{hh}^{(q)}$, causing the thermal width at $T\neq$ 0, are assumed to be equal to ${\cal F}_2$. (See, e.g., Sec. II B of Ref. \cite{PDM2} for the detailed discussion on the justification of these assumptions). The unperturbed energy omega $\omega_q$ and the parameter  ${\cal F}_1$ are chosen to reproduce the experimental value for the energy (around 15 MeV) and the width (around 4 MeV) for the GDR in $^{88}$Mo at $T=$ 0 and $J=$ 0. 
 The parameter ${\cal F}_2$ is chosen so that the GDR energy, which is obtained as the solution of the equation $\omega - \omega_q - P_q(\omega) = 0$ at $J=$ 0, does not change significantly as $T$ varies. The selected values of ${\cal F}_1 = 0.071$ MeV and ${\cal F}_2 = 0.163$ MeV with $E_{GDR}= $ 15 MeV for GDR in $^{88}$Mo are then kept unchanged throughout the calculations as $T$ and $J$ vary. A value $\varepsilon =$ 0.5 MeV is adopted for the smoothing parameter  in Eq. (\ref{gamma1}), which mimics the effect of the escape width caused by coupling to the continuum. 
\subsection{Average temperatures and angular momenta}
\begin{table}
\begin{center}
    \caption
    {Beam and excitation energies, maximal and average temperatures, maximal, highest, and average angular momenta in the fusion-evaporation reaction $^{48}$Ti + $^{48}$Ca $\rightarrow ^{88}$Mo$^{*}$. \label{table1}}
    \vspace{2mm}
\begin{tabular}{|c|c|c|c|c|c|c|c|}
\hline
\hline 
{~$E_{b}$~(MeV)~} & 
{~$E^*$~(MeV)~}&~${T}_{max}$~(MeV)~&
~$\overline{T}$~(MeV)~&
{~${J}_{max}$~($\hbar$)~}&
{~$\sigma$~($\hbar$)~}&
{~${J}_{high}$~($\hbar$)~}&
{~$\overline{J}$~($\hbar$)~}\\
 \hline 
150 & 58 & 1.95 & 1.46 & 32 & 2 & 42 & 22 \\
170 & 67 & 2.19 & 1.61 & 44 & 2 & 52 & 29\\
200 & 80 & 2.49& 1.73 &  55 & 1 & 60 & 37\\
300 & 126 & 3.06& 2.04 & 57 & 2 & 67 & 38\\
400 & 172 & 3.71& 2.43 &  57 & 3 & 72 & 39\\
450 & 194 & 3.92& 2.56 &  57 & 3 & 72 & 39\\
500 & 217 & 4.34& 2.87 &  57 & 3 & 72 & 39\\
600 & 262 & 4.68& 3.06 &  57 & 4 & 77 & 38\\
700 & 308 & 5.12& 3.37 &  56 & 5 & 81 & 38\\
\hline
\hline
\end{tabular}
\end{center}
\end{table}
Shown in Figs. \ref{Tdis} and \ref{Jdis} are the probability distributions $p_T(T_i)$ for temperature and $p_J(J_i)$ for angular momentum as functions of temperature $T$ and angular momentum $J$, respectively. They are obtained by using the  {\scriptsize GEMINI++} code~\cite{Gem_Char, Gem_Cie}, which generates the statistical decays for the recent fusion-evaporation reaction $^{48}$Ti $+$ $^{40}$Ca, and produces the compound nucleus $^{88}$Mo$^{*}$ at 9 values of beam energy $E_b=$ 150, 170, 200, 300, 400, 450, 500, 600, and 700 MeV. The calculations take into account the competition owing to fission under the assumption that the angular momentum of the compound nucleus $^{88}$Mo$^{*}$ is preserved, that is not affected by the decay paths over all daughter nuclei. This assumption is justified by the fact that the GDR energies and widths for the molybdenum isotopes are essentially the same at similar values of $T$ and $J$~\cite{Schiller}. The excitation energy $E^{*}$ of the compound nucleus in a complete fusion is obtained from the beam energy by using the relation $E^{*} = E_{cm} + Q$, where $E_{cm}$ is the total kinetic energy in the center of mass system and the reaction Q value is equal to the sum of projectile and target masses minus the mass of the compound nucleus. The values of excitation energy $E^{*}$ that correspond to these beam energies are deduced as $E^{*} =$ 58, 67, 80, 126, 172, 194, 217, 262, and 308 MeV and listed in the column 2 of Table \ref{table1}. The deduced values of temperature correspond to the evaporation of the daughter nuclei transmitted by the high-energy $\gamma$-rays, that is to the nuclei upon which the GDR is built. The value $J_{max}$ of angular momentum, at which fission starts to set in, is found from the fit of the angular momentum distribution by using the formula $p'_J(J_i)={\cal N}(2J_i+1)/\{1+\exp[(J_i-J_{max})/\sigma]\}$, where $\sigma$ is the diffuseness of the distribution, and ${\cal N}=\sum_ip_J(J_i)/\sum_ip'_J(J_i)$ is the normalization factor.
The value $J_{high}= J_{max} + 5\sigma$ is defined as the highest value of $J_i$, starting from which the high-$J$ tail in the angular momentum distribution at $J>J_{high}$ becomes negligible. Indeed, by using the values $J_{max}$ and $\sigma$ at the excitation energies $E^{*}$ shown in Table \ref{table1}, we found that the sum of $p_J(J)$ within the interval $J_{max} + 5\sigma \leq J \leq$ 100$\hbar$ does not exceed 0.1$\%$ of the total sum of $p_J(J)$  within 0 $\leq J \leq$ 100$\hbar$. Figures. \ref{Tdis} and \ref{Jdis} as well as Table \ref{table1} clearly show that the maximal temperature $T_{max}$ and  the highest angular momentum $J_{high}$ increase with excitation energy $E^{*}$, and reach the values as high as $T_{max}=$ 5.12 MeV and $J_{high}=$ 81$\hbar$ (See the columns 3 and 7 of Table \ref{table1}). However, the average values of temperatures and angular momenta are actually much lower (See the  columns 4 and 8 of Table \ref{table1}). It is also worth noticing that the maximal angular momentum $J_{max}$ increases first with excitation energy $E^*$ up to $E^{*}=$ 172 MeV, where it saturates, because of fission competition, at the value of 56 -- 57$\hbar$.  Consequently the average angular momentum also reaches the maximum equal to 38 -- 39$\hbar$ at $E^{*}\geq$ 172 -- 217 MeV.

\subsection{Average GDR strength functions}
    \begin{figure}
       \includegraphics[width=15cm]{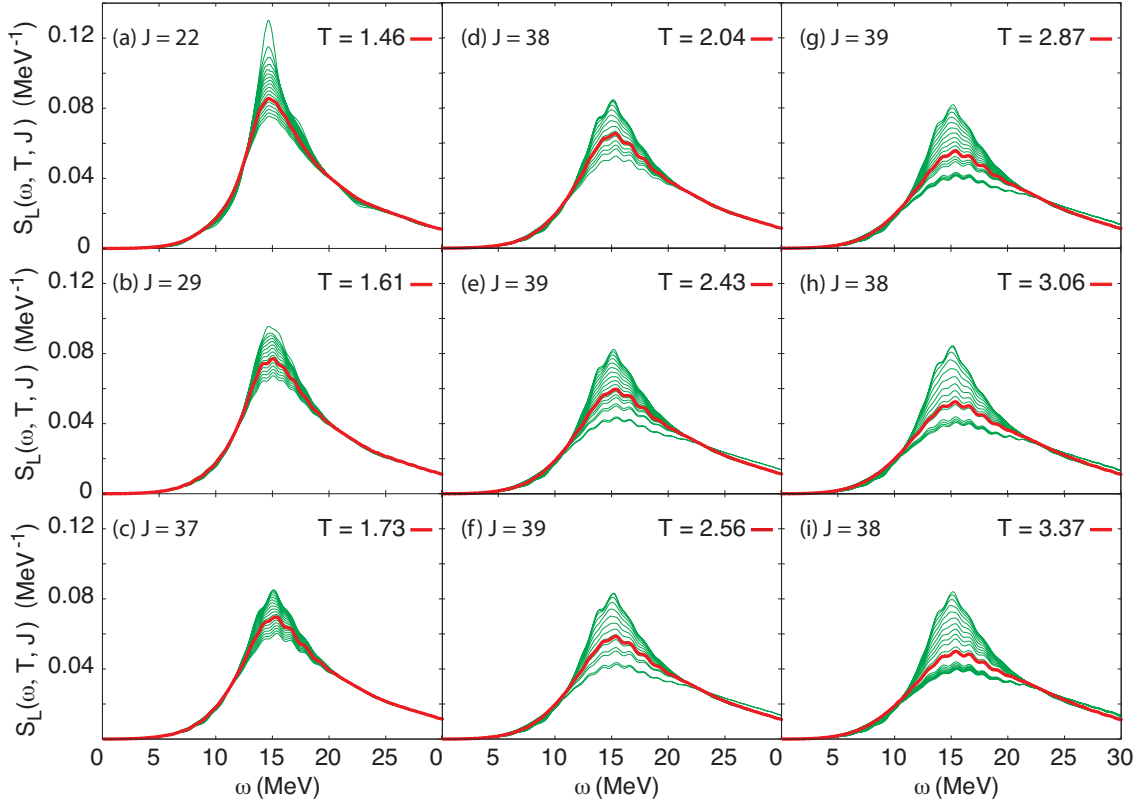}
       \caption{(Color online) GDR strength function $S_L(\omega, T, J)$ for $^{88}$Mo at different values $T = T_i$ taken from the temperature probability distribution $p_T(T_i)$ in Fig. \ref{Tdis} [(green) thin lines with a lower maximum at a higher $T_i$] and at $J = \overline{J}$ (in $\hbar$). The (red) thick solid lines denote $S_L(\omega,\overline{T}, J)$ at $T = \overline{T}$ (in MeV). 
   \label{sMoTJ}}
    \end{figure}
\begin{figure}
       \includegraphics[width=15cm]{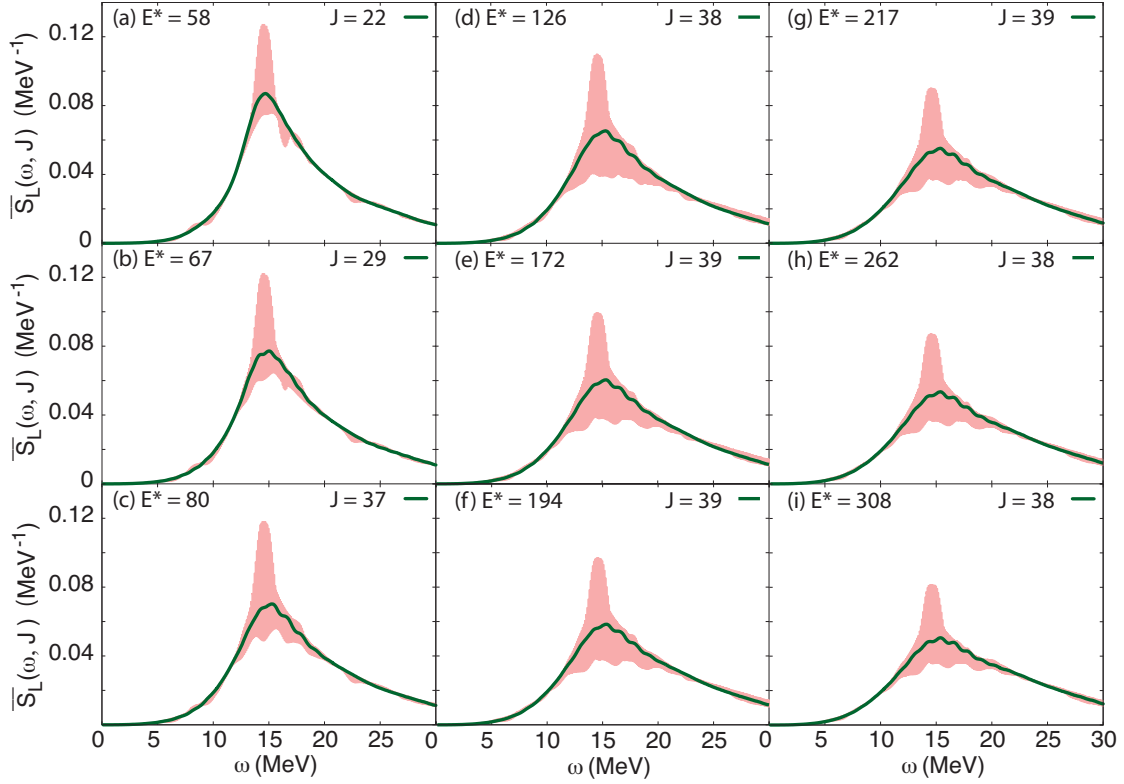}
       \caption{(Color online) GDR average strength function $\overline{S}_L(\omega, J)$ for $^{88}$Mo at different values $J = \overline{J}$ taken from the angular-momentum probability distribution $p_J(J_i)$ in Fig. \ref{Jdis} [(pink) shaded areas]. The (green) solid lines denote $\overline{S}_L(\omega, \overline{J})$ at the values $J=\overline{J}$ (in $\hbar$) shown in the panels. 
          \label{sMoJ}}
    \end{figure}

Displayed in Fig. \ref{sMoTJ}  are the GDR strength functions $S_L(\omega,T, J)$, calculated from Eq. (\ref{SL}) at at various temperatures $T$ and $J$ equal to $\overline{J}$ that correspond to the excitation energies $E^{*}$ listed in Table \ref{table1}. The temperatures $T$ are those $T_i$ at which the temperature probability distribution $p_T(T_i)$ in Fig. \ref{Tdis} is obtained and also $T =\overline{T}$ from Table \ref{table1}. These strength functions are those obtained at given values of $T$ and $J$ at each step of the statistical decay. They are not the actual strength functions, which are generated by averaging over all the cascades. An illustration of a partial averaging is shown in Figs.  \ref{sMoJ}, where the strength functions $\overline{S}_L(\omega, J)$, obtained from Eq. (\ref{averS}) by averaging over the temperature probability distribution $p_T(T_i)$, are shown at $J$ running in step of 1$\hbar$ from 0 up to $J_{max}$.  One can see that both temperature angular momentum variations in the broad regions between 0$\leq T\leq$ 5.21 MeV and 34 $\leq J\leq$ 85$\hbar$ cause large changes in the individual GDR line shapes. 
    \begin{figure}
       \includegraphics[width=14cm]{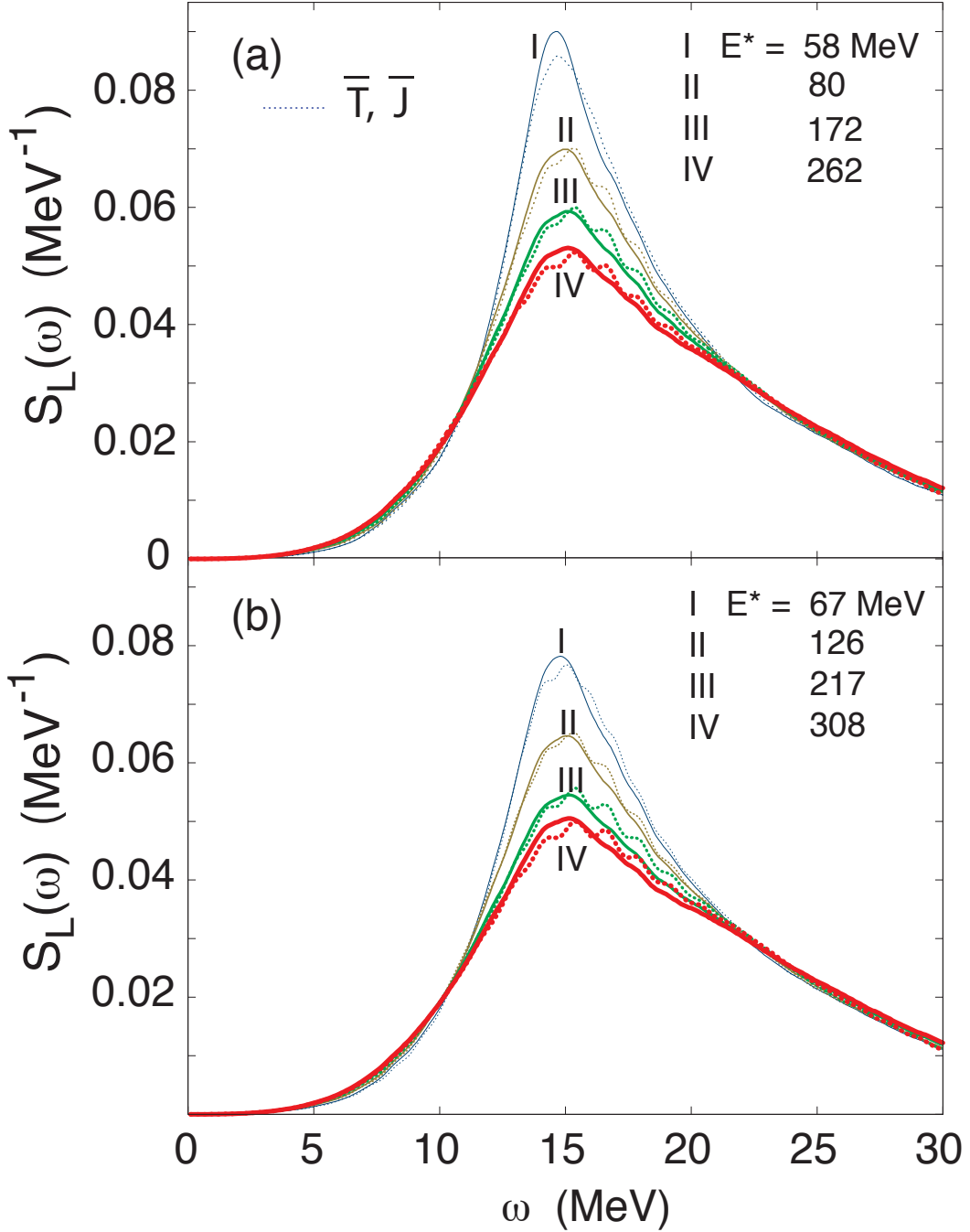}
       \caption{(Color online) GDR average strength function $S_L(\omega, E^{*})$ for $^{88}$Mo at different excitation energies $E^{*}$ (as specified in the panels) obtained by using the temperature probability distribution $p_T(T_i)$ and angular momentum probability distributions $p_J(J_i)$. The dotted lines are the strength functions $S_L(\omega, T, J)$ obtained at the corresponding $T =\overline{T}$ and $J=\overline{J}$ (as in Table \ref{table1}). 
   \label{savMo}}
    \end{figure}
    \begin{figure}
       \includegraphics[width=12cm]{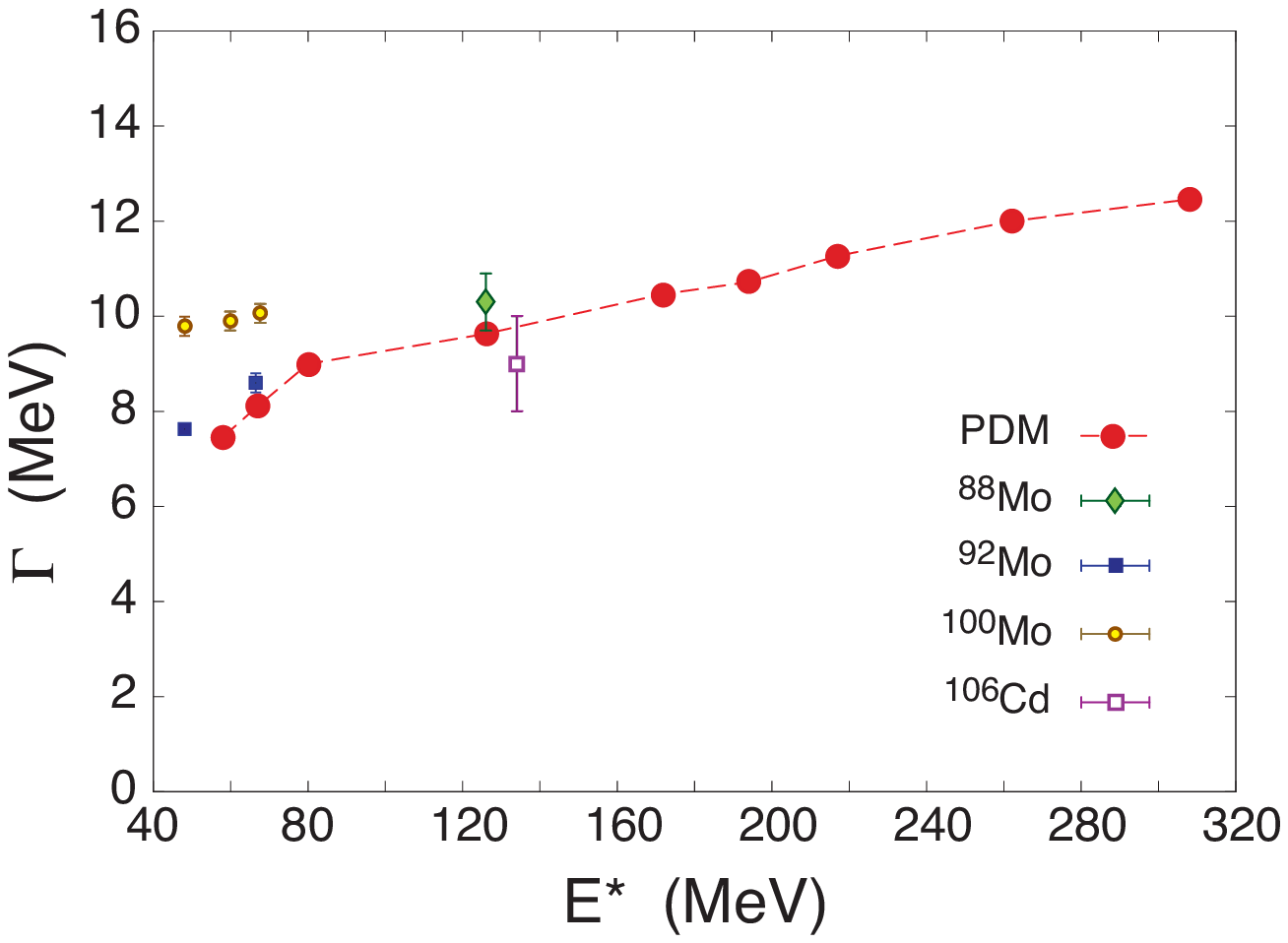}
       \caption{(Color online) GDR width in the fusion-evaporation reaction $^{48}$Ti + $^{48}$Ca $\rightarrow ^{88}$Mo$^{*}$ as a function of $E^{*}$.  The (red) full circles are PDM predictions, connected with the dashed line to guide the eye. The experimental data for $^{88,92,100}$Mo and $^{106}$Cd are taken from Refs. \cite{Mo88,Mo88_2,Kic,Luc}. \label{widthMo}}
    \end{figure}

The central results of the present paper are displayed in Fig. \ref{savMo}, where the GDR average strength function $\overline{S}_L(\omega, E^{*})$ (\ref{averS}) for $^{88}$Mo is compared with the strength function $S_L(\omega,\overline{T},\overline{J})$ obtained at $(T, J) = (\overline{T}, \overline{J})$ at various $E^{*}$. It is seen from this figure that the GDR width increases with $E^{*}$ despite the fact that the value $\overline{J}$ of the average angular momentum actually decreases with increasing $E^{*}>$ 217 MeV. The most important feature is that the GDR line shape obtained by averaging over the temperature and angular-momentum probability distributions is very similar to that obtained at the average values ($\overline{T}, \overline{J}$) of temperature and angular momentum in these fusion-evaporation reactions, where the compound nucleus $^{88}$Mo is formed and produces the GDR in its subsequent statistical decays. In fact, except for a slight difference around the GDR peak, the average strength function  $\overline{S}_L(\omega, E^{*})$ practically coincides with $S_L(\omega, \overline{T}, \overline{J})$ with the same FWHM.

    \begin{figure}
       \includegraphics[width=11cm]{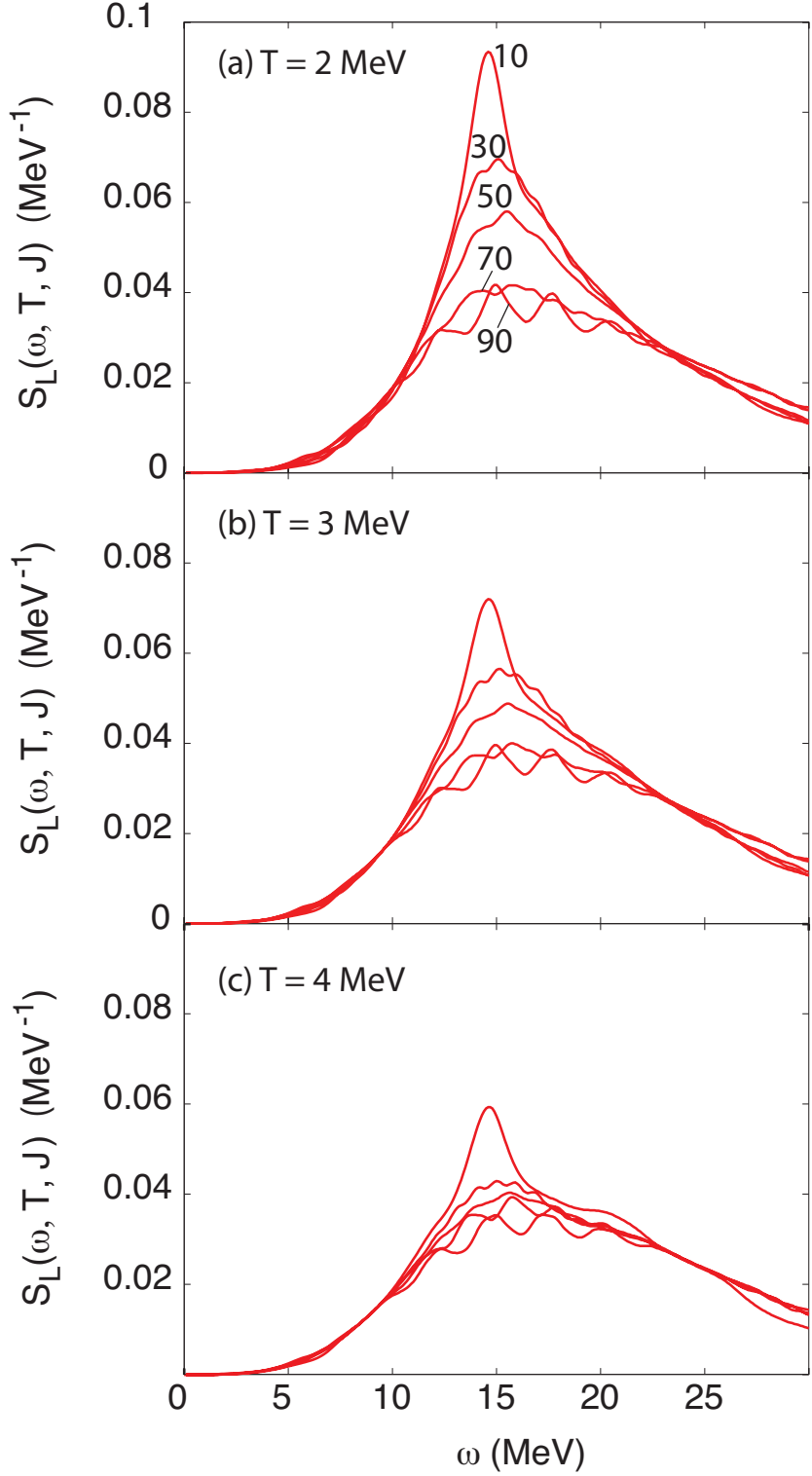}
       \caption{(Color online) GDR strength functions $S_L(\omega, T, J)$ for $^{88}$Mo at several temperatures $T=$ 2, 3, and 4 MeV and angular momenta $J=$ 10, 30, 50, 70 and 90$\hbar$ shown as the numbers at the corresponding lines. \label{sMotest}}
    \end{figure}
    \begin{figure}
       \includegraphics[width=12cm]{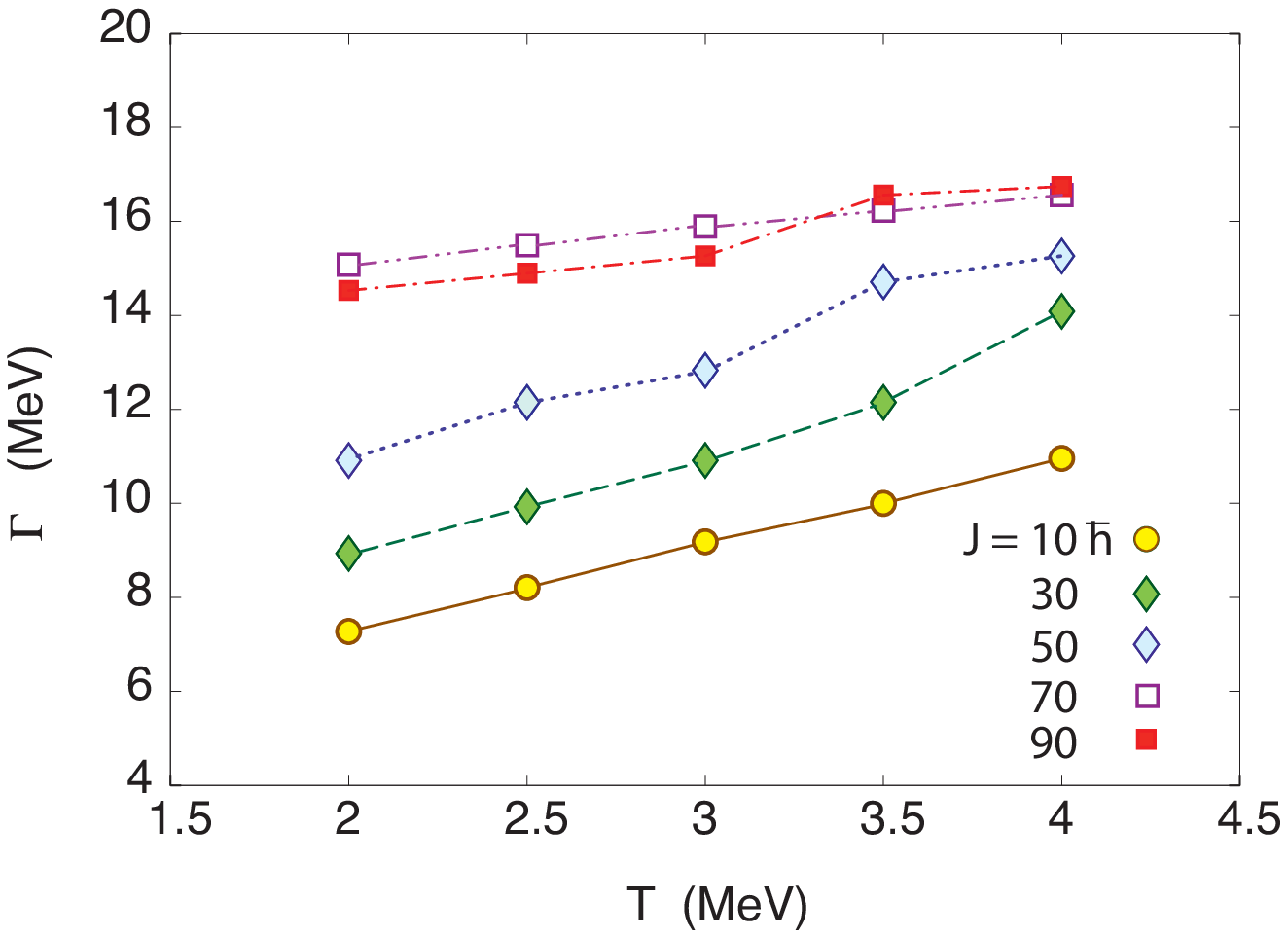}
       \caption{(Color online) GDR width in $^{88}$Mo vs 2 $\leq T \leq$ 4 MeV at 10 $\leq J\leq$ 90$\hbar$. The lines are drawn to guide the eye.   \label{widthMotest}}
    \end{figure}
The GDR width obtained within the PDM for $^{88}$Mo is plotted against $E^{*}$ in Fig. \ref{widthMo} in comparison with the available GDR experimental widths for molybdenum isotopes. The data point for $^{88}$Mo is taken from Refs. \cite{Mo88, Mo88_2}, and those for $^{92,100}$Mo are from Ref. \cite{Kic}. Also shown is the GDR experimental width for $^{106}$Cd~\cite{Luc}, which is located approximately at the same $E_{GDR}$ at that for $^{88}$Mo. The GDR experimental line shapes in $^{92,100}$Mo have been obtained by fitting the $\gamma$-ray spectra with two Lorentzians located at $E_1\sim$ 14.5 -- 15 MeV and $E_2 \simeq$ 19 MeV, showing that these isotopes are well deformed nuclei. This explains why the FWHM of the GDR in $^{92}$Mo and especially $^{100}$Mo, where $E_2/E_1$ reaches 1.28 -- 1.58, are larger than the corresponding PDM predictions for the GDR width in spherical $^{88}$Mo. In general At $E^{*}\leq$ 80 MeV the increase in the width is rather strong, but at $E^{*}>$ 80 MeV the width increase is weaker because of the saturation of $J_{max}$. To see if the width saturation at high excitation energy is caused by the saturation of angular momentum, a test is made to examine the competition between the temperature and angular momentum effects on the width increase as $T$ varies from 2 to 4 MeV in step of 0.5 MeV and $J$ is allowed to increase from 10 to 90$\hbar$ in step of 20$\hbar$. The GDR strength functions $S_L(\omega, T, J)$ obtained at $T=$ 2, 3, and 4 and at all $J$ in use are displayed  in Fig. \ref{sMotest}, whereas the GDR widths are plotted in Fig. \ref{widthMotest} versus $T$ at all values of $J$ in use. These figures clearly show a significant contribution of the angular momentum effect to the width increase at a lower $T$ ($T=$ 2 MeV), whereas at high $T=$ 4 MeV the width obviously goes to a saturation at $J\geq$ 50$\hbar$. Moreover, at larger $J\geq$ 70$\hbar$, a width saturation starts to take place at any $T$. This feature is in qualitative agreement with the result obtained previously in Ref. \cite{PDM3}.
\section{Conclusions}
In the present paper the PDM is employed to calculate the strength functions for the GDR in the statistical decays after the fusion-evaporation reaction$^{48}$Ti $+$ $^{40}$Ca, which produces the compound nucleus $^{88}$Mo$^{*}$ at various excitation energies $E^{*}$.
The calculations use the empirical probability distributions for temperature and angular momentum, which are generated by the {\scriptsize GEMINI++} statistical code to produce the GDR average strength functions $\overline{S}_L(\omega, E^{*})$ as well as the average temperature $\overline{T}$ and average angular momentum $\overline{J}$ at each energy $E^{*}$. 

The calculations show that, while the GDR width increases with $E^{*}$, it approaches a saturation at high $T=$ 4 MeV when the angular momentum $J$ reaches the value larger than 50$\hbar$. At a larger $J\geq$ 70$\hbar$, the width saturation shows up at any $T$. The most important observation in the present paper in that the GDR strength function $\overline{S}_L(\omega, E^{*})$ obtained by averaging the individual strength functions $S_L(\omega,T,J)$ over the empirical temperature and angular-momentum probability distributions turns out to be almost identical to $S_L(\omega,\overline{T},\overline{J})$ calculated at the average values $\overline{T}$ and $\overline{J}$ of temperature and angular momentum. This conclusion has a practical importance in the comparison between theory and experiment since, once $\overline{T}$ are $\overline{J}$ are known, one may compare the theoretical prediction for the individual strength function $S_L(\omega,T,J)$ and its width, obtained at $\overline{T}$ and $\overline{J}$, with the data, without the need of generating and averaging the strength functions over the whole temperature and angular momentum distributions. For the direct comparison with experimental data, the changes in the original angular momentum distribution of the compound nucleus during the evaporation process have also to be taken into account. This will be carried out in the forthcoming paper \cite{Mo88_2}.
\acknowledgments
The numerical calculations were carried out using the {\scriptsize FORTRAN IMSL}
Library by Visual Numerics on the RIKEN Integrated Cluster of Clusters (RICC) system. 
M.C., M.K. and A.M. acknowledge the partial support from the  Polish National Center of Science (grants No. 2011/03/B/ST2/01894 and N~N202 486339).

\end{document}